\documentclass{article}
\usepackage[affil-it]{authblk}
\usepackage[utf8]{inputenc}
\usepackage{graphicx}
\usepackage{grffile}
\usepackage{amsmath}
\usepackage{amssymb}
\usepackage{hyperref}
\usepackage{subcaption}
\usepackage{floatpag}
\usepackage{makeidx}
\usepackage{amsmath}

\usepackage{amsthm}
\usepackage{amsfonts}
\usepackage{amssymb}
\usepackage{algorithm}
\usepackage{algpseudocode}
\makeindex
\usepackage{accents} 
\DeclareMathAlphabet{\mathpzc}{OT1}{pzc}{m}{it} 
\numberwithin{equation}{section}

\newcommand{\mmis}[2]{\ensuremath{\underaccent{#2}{\mbox{#1}\,\,}}}
\renewcommand{\max}[1]{\mmis{max}{#1}}
\renewcommand{\min}[1]{\mmis{min}{#1}}

\renewcommand{\lim}[1]{\mmis{lim}{#1}}

\newcommand{\norm}[1]{\ensuremath{\left|\left|#1\right|\right|}}

\usepackage{mathtools}

\DeclarePairedDelimiter\floor{\lfloor}{\rfloor}
\newcommand{\argmin}[1]{\mmis{argmin}{#1}}
\newcommand{\argmax}[1]{\mmis{argmax}{#1}}

\begin{document}

\markboth{Renato Budinich}{A Region Based Easy Path Wavelet Transform for Sparse Image Representation}

\date{\today}
\title{A Region Based Easy Path Wavelet Transform\\
  For Sparse Image Representation }

\author{Renato Budinich}

\affil{Institut für Numerische und Angewandte Mathematik, Universität Göttingen \\
Lotzestrasse 16-18, 37083 Göttingen, Deutschland\\
r.budinich@math.uni-goettingen.de}

\maketitle

\begin{abstract}
The \emph{Easy Path Wavelet Transform} is an adaptive transform for bivariate functions (in particular natural images) which has been proposed in \cite{epwt}. It provides a sparse representation by finding a path in the domain of the function leveraging the local correlations of the function values. It then applies a one dimensional wavelet transform to the obtained vector, decimates the points and iterates the procedure. The main drawback of such method is the need to store, for each level of the transform, the path which vectorizes the two dimensional data. Here we propose a variation on the method which consists of firstly applying a segmentation procedure to the function domain, partitioning it into regions where the variation in the function values is low; in a second step, inside each such region, a path is found in some deterministic way, i.e. not data-dependent. This circumvents the need to store the paths at each level, while still obtaining good quality lossy compression. This method is particularly well suited to encode a Region of Interest in the image with different quality than the rest of the image.
\end{abstract}

\section{Introduction}
\label{sec:orgheadline2}

In recent years there has been a big interest in using wavelet-like transforms for natural image compression and denoising. One wishes to find an appropriate transform such that the most important information of the image is concentrated in few coefficients; by then thresholding coefficients and keeping only the largest ones, one hopes to obtain a good quality approximation of the original image. However simply using a two-dimensional tensor wavelet transform doesn't yield good results, mainly because edges are poorly preserved. This is due to the support of the basis elements not being adapted to the directional geometric properties present in natural images.

\noindent Therefore transforms such as curvelets \cite{candes2006fast,candes2004new} and shearlets \cite{guo2004wavelets,guo2007optimally} have been developed, which are highly redundant frames with strong anisotropic selectivity. However these are non-adaptive frames which loose their near optimal approximation properties if strong hypotheses are dropped on the edges in the image (namely piecewise \(C^2\)). 

\noindent Instead of choosing an a priori frame to approximate the image, one can instead choose to adapt the frame elements to the particular image. Many different approaches to this concept have been studied in the recent years, such as bandlets \cite{le2005bandelet}, grouplets \cite{mallat2009geometrical} and dictionary learning methods \cite{aharon2006img,rubinstein2010double}. For a review of adaptive image representations see \cite{peyre2011review}.

\noindent In \cite{epwt} a new adaptive transform for images termed \emph{Easy Path Wavelet Transform} (EPWT) was introduced. In this method, a path is found among the points of the image that leverages local correlation in the gray values so as to produce a one dimensional signal with low entropy. At every level of the transform such a path is found and then a one dimensional wavelet transform is applied. The image quality obtained with this method is very good when compared to other methods; the main drawback comes from the need to store the paths for each level, which are needed during decoding. In \cite{plonka2012optimally} and \cite{plonka2013optimal} it was shown that, with a suitable choiche of the paths, the \(N\)-term approximation given by the EPWT is optimal for piecewise-Hölder functions. In \cite{plonka2011new} the EPWT was used as part of a hybrid method for Image Approximation while in \cite{hp12} for denoising of scattered data.

Here we propose a variation on the original EPWT method, which we call the \emph{Region Based Easy Path Wavelet Transform} (RBEPWT) method. The objective is to reduce the adaptivity storage cost by not requiring to store the paths like in the EPWT. In order to achieve this, a segmentation method is applied to the image in a first step, in order to partition the image into areas of low variation of the gray values. Then, for each region, a path is found in some kind of canonical manner: the path depends only on the geometrical shape of the region's border and not on the gray-values of the inner pixels. In this way in the final encoding one needs to store only the regions obtained from the initial segmentation step and the wavelet coefficients; the paths, being a deterministic function of the region's shape, can be recomputed on-the-fly during decoding, and thus need not be stored.

\noindent The quality of the lossy compression obtained from decoding a hard-thresholded set of coefficients heavily depends on the initial segmentation. In this regard we require a segmentation that finds regions where the local variance in the gray-values of the pixels is low - we do not care if the  identified regions correspond to semantic areas in the image (which is the objective of many segmentation methods used in computer vision), but we wish instead to not have big jumps in the gray-values in the regions. We thus need a segmentation algorithm that essentially does a clustering: if we think of the image as a set of points in a three-dimensional space, where the first two coordinates are the indexes of the pixel and the third its gray-value, we wish for the segmentation to identify the main clusters of such points. 

\noindent In all our preliminary numerical tests we used the Felzenszwalb-Huttenlocher algorithm proposed in \cite{felz-hutt}, because of its excellent quality to computation speed ratio. In the future we would like to further investigate this important first step in our method, and develop an ad hoc procedure to find regions that are already somehow optimal for the further steps of the RBEPWT transform. We will comment more on this in Section \ref{orgtarget1}.

The outline of this paper is as follows. In Section \ref{orgtarget2} we introduce a common framework for the EPWT and RBEPWT. In Sections \ref{orgtarget3} and \ref{orgtarget4} we specify the particular details of the EPWT and RBEPWT respectively when viewed in this framework; in particular in Section \ref{orgtarget1} we comment on the desirable characteristics of a segmentation method for the RBEPWT. Finally in Section \ref{orgtarget5} we present the results of various numerical experiments.  

\section{Easy-path wavelet transform methods}
\label{sec:orgheadline9}
\label{orgtarget2}
Both the original EPWT method and the here proposed RBEPWT conform to a common general framework.  Let \(f:I \rightarrow \{0,\ldots,255\}\) be the input image, where \(I = \{ (i,j) \,\, | \,\, 0 \leq i,j \leq N-1 \}\) is the set of indexes and \(N = 2^k\) for some \(k \in \mathbb{N}\); assume \(I\) is ordered in some canonical manner, for example using the lexicographical ordering. Let \(L\in \mathbb{N}\) be the number of levels of the transform; \(2^{L} \leq N^{2}\) must hold. Choose a set of low-pass and high-pass 1-dimensional wavelet synthesis and reconstruction filters; following notation from \cite{mallat2008wavelet} we will call them \(h,g,\bar{h},\bar{g}\).

\noindent Define \(I^L := I\) and \(f^L := f\). The encoding for the first level consists of:
\begin{enumerate}
\item Finding a \emph{path} in \(I^L\); this can be thought equivalently as a function 
\begin{align}
\label{eq:path}
p^L:\{0,1,\ldots,|I|-1\} \rightarrow I^L
\end{align}
or as a permutation  of the elements of \(I^L\).  How this path is found is the central point of such methods, and what differentiates the EPWT from the RBEPWT; we will comment more on this later. We can then define \(\tilde{f}^L : I^L \rightarrow \{0,\ldots,255\}\) by \(\tilde{f}^L := f^L \circ p^L\) as the 1-dimensional signal obtained from sampling the image along the path.
\item Apply one level of a periodic 1-dimensional discrete wavelet transform to \(\tilde{f}^L\), obtaining approximation and detail coefficients \(a^L\) and \(d^L\) respectively:
\begin{align}\label{dwt:lev1}
\begin{split}
a^L(k) &:= \tilde{f}^L \star \bar{h} (2k) \\
d^L(k) &:= \tilde{f}^L \star \bar{g} (2k), \quad k=0,\ldots,\frac{|I^L|}{2}-1
\end{split}
\end{align}
where \(\star\) denotes the discrete convolution.
\item Define \(I^{L-1}:= \{ p^L(2k) \text{ s.t. } k=0,\ldots,\frac{|I^L|}{2}-1 \}\); it is the subset of \(I^L\) obtained by taking only the coordinates corresponding to even indexes in the path \(p^L\), i.e. it is \(I^L\) decimated by a factor of \(2\), following the order induced by \(p^L\). Define \(f^{L-1} : I^{L-1} \rightarrow \{0,\ldots,255\}\) by \(f^{L-1}(p^L(2k)) =   a^L(k)\).
\end{enumerate}
Now, \(I^{L-1}\) and \(f^{L-1}\) are a new set of indexes and a vector of values respectively, with half the points of \(I^L\). One can thus iterate the \(3\) steps of encoding, at each level halving the number of points. 

\noindent In summary the encoding steps of the transform consist of:
\begin{enumerate}
\item permuting the order of the points (i.e. find a path),
\item applying a discrete wavelet transform,
\item storing the detail vector and use the approximation vector as values for the new level, which is obtained by down-sampling the points along the found path.
\end{enumerate}

\noindent Since we are interested in lossy compression, typically after the encoding a thresholding procedure will be used on the coefficients. Following the usual assumption that the coefficients with smallest absolute value in the encoded image are the least significant ones, we simply keep the \(n\) largest coefficients (in absolute value) and set the others to \(0\).

\noindent For decoding one needs the approximation vector for the lowest level, all the wavelet detail vectors and the permutations for each level. Then the decoding procedure consists simply of
\begin{enumerate}
\item applying the inverse wavelet transform for that level,
\item applying the inverse permutation for that level.
\end{enumerate}

Note that, since we are only interested in the final result of the decoding (where the spatial disposition of the points is given by the canonical ordering chosen in advance), we do not need to know, for the intermediate levels,  which value in the vector corresponds to which point in space. In other words, while during encoding we down-sample both the points in space and the vectors of values associated to them, in the decoding phase we can operate on the vectors only, upsampling and permuting them. The spatial information is needed only during encoding in order to find the path, i.e. the order according to which we vectorize the 2-dimensional data; during decoding this order (given by the inverse permutations) is already known.

\subsection{EPWT}
\label{sec:orgheadline3}
\label{orgtarget3}
In the EPWT method the path at level \(l\) starts from a canonical point (for example \((0,0)\)) and at each step, among the closest points that are still avaiable (i.e. aren't already part of the path), it makes a greedy choice: the point that gives the least difference in absolute value is chosen
\begin{align}\label{epwt:path}
&\begin{cases}
p^{l}(0) &:= (0,0) \\
p^{l}(k+1) & \in \argmin{(i,j) \in I^{l}\cap B^{\circ}_{\tilde{h}}(p^{l}(k))}  |f^{l}(i,j) - f^{l}(p^{l}(k))|
\end{cases}\\
&\mbox{where } \tilde{h} := \min{}\{h \in \mathbb{N}, \,\, B^{\circ}_h(p^{l}(k)) \not\subseteq \cup_{m=1}^{k-1}p^{l}(m), \,\, B^{\circ}_h(p^{l}(k)) \cap I^{l} \neq \emptyset \} \,\,.  \nonumber 
\end{align}
Here \(B^{\circ}_h((i,j))\) denotes the punctured ball\footnote{we use either the \(\max{}\) distance, defined by \(d((i,j),(k,l)) = \max{} \{|i-k|,|j-l|\}\), or the usual Euclidean distance} centered in point \((i,j)\) of radius \(h\) and \(\tilde{h}\) is the minimum required radius to obtain a neighborhood of point \((i,j)\) that contains points in \(I^l\) that have not yet been taken as previous points of the path. If the \(\argmin{}\) set contains more than one element, the choice can be done so as to minimize the direction change in the path.

\noindent In order to have access to the inverse permutations during encoding, the paths for each level have to be stored, alongside the wavelet detail and approximation coefficients. In \cite{epwt} it was shown that, with the same number of coefficients, the EPWT greatly outperforms the classical 2-dimensional wavelet transform. However the storage cost is strongly affected by the need to store all the permutations; this is the point we wish to address here with the new RBEPWT method.

\subsection{RBEPWT}
\label{sec:orgheadline7}
\label{orgtarget4}
In the region based EPWT, we first apply a segmentation method to the image, in order to obtain a subdivision of the original image into \emph{regions} (subsets of points). The rationale is that if we have regions where the variation in gray-values is small, then it is not so important which path is taken inside that region. We can then have a canonical path-finding procedure which does not depend on the gray-values. In this way the final encoding consists of the wavelet detail, the approximation coefficients and the segmentation. We need not to store all the paths, since these depend only on the segmentation and not on the pixel grayvalues and thus they can be recomputed on-the-fly during decoding.

\noindent We will comment in Section \ref{orgtarget1} on the properties the segmentation method should have. For now we suppose that the segmentation step identifies regions \({R_0,R_1,\ldots, R_{r-1} \in \mathcal{P}(I)}\) where \(I\) is the set of indexes in the image and \(\mathcal{P}(I)\) denotes the set of sets of \(I\). The regions are given in the form of a \emph{label image} \(\Lambda \in \{0,1,\ldots,r-1\}^{N\times N}\), obtained by filling region \(R_h\) with the value \(h\), i.e. such that \((i,j) \in R_h \Leftrightarrow \Lambda_{i,j} = h\). 

\noindent Suppose now we have a function \(\Pi\) that associates to any set of points a Hamiltonian path (in the complete graph generated by these points). In other words, for any region \(R\) we wish for \(\Pi(R)\) to be a bijection from \(\{0,1,\ldots, |R|-1\}\) to \(R\). Later we will present two examples of such functions, the easy-path and the grad-path.

\noindent Call \(R_k^L := R_k\) for all \(k=0,\ldots,r-1\) and define the region collection at the highest level as \(\mathcal{R}^L := \{ R_0^L,R_1^L,\ldots, R_{r-1}^L\}\). Define for each region \(k\) the path in it as  \(\rho_k^L := \Pi(R_k^L)\). By gluing all these paths together we obtain 
\begin{align*}
p^L := \rho_0^L \cup \rho_1^{L} \cup \ldots \cup \rho_{r-1}^L,
\end{align*}
which is a permutation of the elements of \(I^L\). At this point a global path is defined, and we can proceed just as described in Section \ref{orgtarget2} for the general framework: define \(\tilde{f}^L := f^L \circ p^L\), compute \(a^L\) and \(d^L\) through discrete convolution with the two filters, define \(I^{L-1}\) and \(f^{L-1}\). Additionally now we have to define the new region collection  \(\mathcal{R}^{L-1} := \{R_0^{L-1},R_1^{L-1},\ldots,R_{r-1}^{L-1}\}\), where  \(R_k^{L-1} = R_k^L \cap I^{L-1}\). An equivalent definition would be:
\begin{align}\label{rbepwt:regions}
\begin{split}
R_0^{L-1} &= \rho_0^L(\mbox{even numbers})\\
R_k^{L-1} &= 
\begin{cases}
\rho_k^L (\mbox{even numbers}) & \mbox{if }  |R_{k-1}^{L-1}|  \mbox{ is even} \veebar \,\, R_{k-1}^{L-1} = \rho_{k-1}^L(\mbox{odd numbers}) \\
\rho_k^L (\mbox{odd numbers}) & \mbox{otherwise}\, ,
\end{cases}
\end{split}
\end{align}
where here \(\veebar\) denotes the exclusive \texttt{OR}, i.e. it evaluates to true when only one of the two conditions is true. This procedure can be repeated analogously for each level. As already mentioned, the final encoding of the image will consist of the segmentation information, all the wavelet detail coefficients and the approximation coefficients for the lowest level.

\noindent For decoding, since the permutations here aren't stored, we first need to recompute them; to do so we simply need to apply the whole encoding procedure ignoring the pixel values but not the segmentation information, which has been stored and is thus available. 

\noindent By what has been said it is clear that for our method, the path finding procedure \(\Pi\) must have the following characteristics:
\begin{enumerate}
\item It must not depend on the points in the region being uniform: from level \(L-1\) onward in fact the points in each region will usually not be uniformly distributed.
\item It must be completely deterministic: this is needed in order to obtain the same paths during encoding and decoding.
\item It must not depend on the gray-values of the image: these aren't available during decoding. An exception can be made, as in the grad-path, if one is willing to store additional information in the final encoding.
\end{enumerate}

\subsubsection{Easy-Path}
\label{sec:orgheadline4}
\label{orgtarget6}
\begin{algorithm}
\caption{Easy-Path algorithm}
\label{algo:easy-path}
\begin{algorithmic}[5]
\Require region \( R \)
\Ensure path \( \pi \)
\State choose starting point \( p \in R \) \Comment{\( p \) is the current point}
\State \( Q = R \setminus p \) \Comment{\( Q \) is the set of avaiable points}
\State \( v = (1,0) \) \Comment{\( v \) is the preferred direction}
\While{\(Q \neq \emptyset \)}
\State \( \bar{h} = \min{}\{h \in \mathbb{N}, \,\, B_h(p) \cap Q \neq \emptyset \}\)
\State \( C = \argmin{\kappa \in B_{\bar{h}}(p)\cap Q} \norm{p-\kappa} \)
\If{\( |C| \geq 2} \)
\State \( C = \argmax{\phi \in C} <\phi-p,v>\) \label{algo1:sp}
\If{\( |C| \geq 2 \)}
\State rotate \( v \) by \( \pi/2 \)
\State \textbf{goto} \ref{algo1:sp}
\EndIf
\EndIf
\State pick \( \psi \in C \) \Comment{there is no choice to be done here: \( C = \{\psi\} \)}
\State append \( \psi \) to \( \pi \)
\State \( v = \psi - p \)
\State \( p = \psi \)
\State remove \( \psi \) from \( Q \)
\EndWhile
\State \Return \( \pi \)
\end{algorithmic}
\end{algorithm}
\begin{figure}[h!]
\begin{subfigure}[b]{0.48\textwidth}
\includegraphics[scale=0.4]{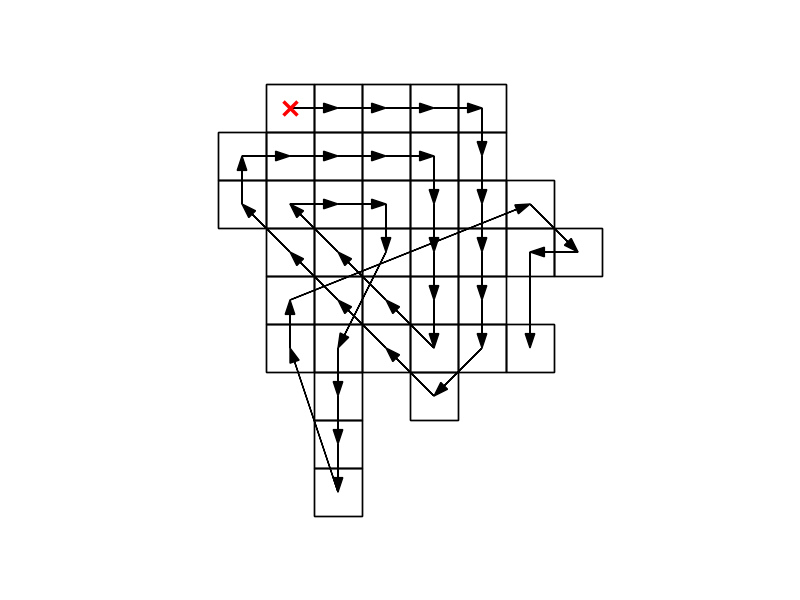}
\end{subfigure} \hspace*{\fill}
\begin{subfigure}[b]{0.48\textwidth}
\includegraphics[scale=0.4]{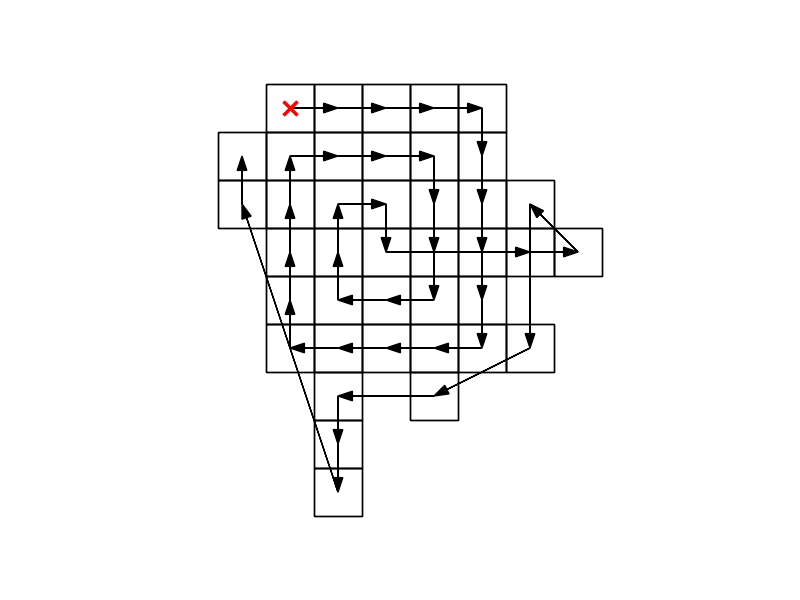}
\end{subfigure}
\caption{Path found by the easy-path procedure with the \( \protect\max{} \) (left) and Euclidean (right) distance\label{fig:regions} }
\end{figure}
Algorithm \ref{algo:easy-path} shows the easy-path procedure to find a path \(\pi\) in a region \(R\): starting from some point (which for example can be chosen using the lexicographical ordering), the algorithm tries always to select the closest avaiable neighbour. If there are more points equally close, it selects the one that would make for the straightest path, the rationale being that a more regular path will lead to a smoother signal (see for example \cite{hp12} for a proof that, when \(f\) is sufficiently smooth, a straighter path gives smaller wavelet coefficients). This is done by computing the scalar product of the increment vector with a preferred direction vector, which at every iteration is updated to be the last increment in the path. If there are \(2\) possible points with the same minimum angle to the preceeding part of the path, then the preferred direction is rotated by \(\pi/2\), making then only one of the two points preferrable.

\noindent See Figure \ref{fig:regions} for an example of a path determined by easy-path. In our tests using the Euclidean distance gave better compression performance, so we used that in all numerical results presented here.
\subsubsection{Grad-path}
\label{sec:orgheadline5}
\begin{algorithm}
\caption{Grad-path algorithm}
\label{algo:grad-path}
\begin{algorithmic}[5]
\Require region \( R \), average gradient \( g \)
\Ensure path \( \pi \)
\State choose starting point \( p \in R \) \Comment{\( p \) is the current point}
\State \( Q = R \setminus p\) \Comment{\( Q \) is the set of avaiable points}
\State \( v =  \) rotate \( g \) by \( \pi/2 \) \Comment{\( v \) is perpendicular to the average gradient}
\While{\(Q \neq \emptyset \)}
\State \( \bar{h} = \min{}\{h \in \mathbb{N}, \,\, B_h(p) \cap Q \neq \emptyset \}\)
\State \( C = \argmin{\kappa \in B_{\bar{h}}(p)\cap Q} \norm{p-\kappa} \)
\If{\( |C| \geq 2} \)
\State \( C = \argmax{\phi \in C} \left|<\phi-p,v>\right|\)  \label{algo2:sp}
\If{\( |C| \geq 2 \)}
\State rotate \( v \) by \( \pi/2 \)
\State \textbf{goto} \ref{algo2:sp}
\EndIf
\EndIf
\State pick \( \psi \in C \) \Comment{there is no choice to be done here: \( C = \{\psi\} \)}
\State append \( \psi \) to \( \pi \)
\State \( w =  \) rotate \( g \) by \( \pi/2 \) 
\State \( v = \argmax{\gamma = -w,w}  <\psi - p, \gamma> \)
\State \( p = \psi \)
\State remove \( \psi \) from \( Q \)
\EndWhile
\State \Return \( \pi \)
\end{algorithmic}
\end{algorithm}

\noindent Another path-finding procedure is the grad-path, shown in Algorithm \ref{algo:grad-path}. It requires previous computation of the average discretized gradient for each region; these vectors have to be stored, contributing to the storage cost of the final encoding. The procedure is very similar to the easy-path: the closest point is always preferred. However the preferred direction is always perpendicular to the average gradient, at each iteration the sign being updated so as to obtain the most regular path. Furthermore taking the absolute value when computing the scalar product (see line \ref{algo2:sp}) means that we always prefer a path that remains as much as possible perpendicular to the average gradient, even if it means a sharper change in the path's direction. Only in case of equally distant points forming equal angles the preferred direction is temporarily rotated.
\begin{figure}[h]
\begin{subfigure}[b]{0.48\textwidth}
\includegraphics[scale=0.37]{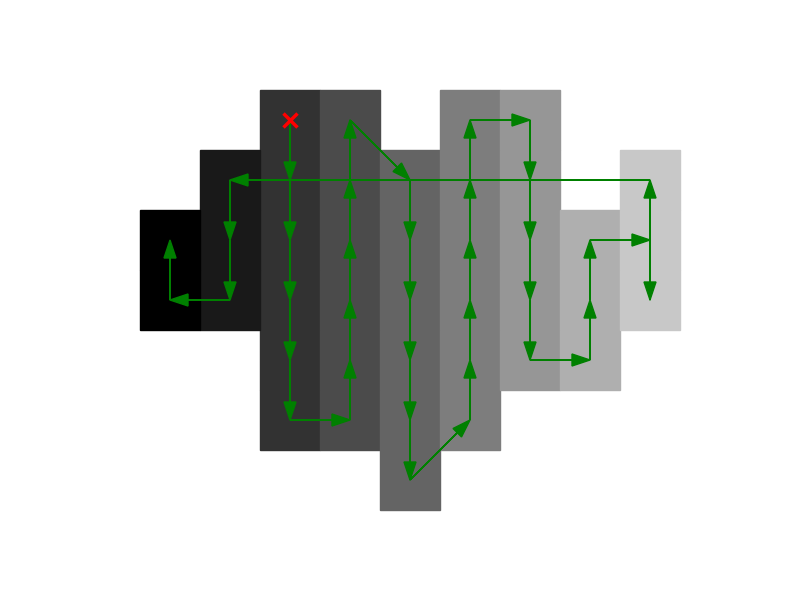}
\end{subfigure} \hspace*{\fill}
\begin{subfigure}[b]{0.48\textwidth}
\includegraphics[scale=0.37]{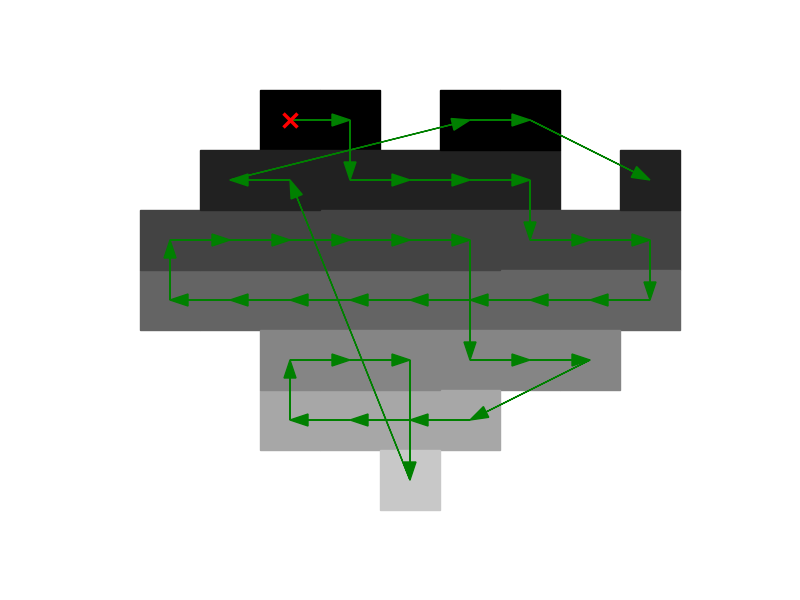}
\end{subfigure}
\caption{Gradpath on same region with different gray-values\label{fig:gradpath} }
\end{figure}

\noindent See Figure \ref{fig:gradpath} for an example of the path generated by the gradpath algorithm. The points forming the region are the same in the two images, but the gray-values generate two different average gradients. 

\noindent In our numerical tests (see Section \ref{orgtarget5}) the grad-path performed only marginally better than the easy-path - this is without taking into account the additional storage cost.
\subsubsection{Notes on the segmentation procedure}
\label{sec:orgheadline6}
\label{orgtarget1}
In order to avoid the need to store the permutation at each level as happens in the EPWT, in the proposed RBEPWT we are finding paths that do not depend on the data, at least not directly\footnote{In this section we are assuming that the easy-path procedure from Section \ref{orgtarget6} is being used}. In fact, the segmentation uses the gray-value information, but once the regions are determined the paths along which we sample are completely determined by the regions and are agnostic of the gray-value data. It is thus clear that the segmentation procedure is of primary importance for our method. 

\noindent Given a segmentation \(\mathcal{R} = \{ R_0, \ldots, R_{r-1} \}\) of the index set \(I\), we can define the \emph{perimeter} of \(\mathcal{R}\) as
\begin{align*}
  P(\mathcal{R}) := \bigg| \Big\{ \big((i,j),(k,l)\big) \text{ s.t. } \norm{(i,j) - (k,l)} = 1 &, \,\, (i,j) \in R_a,  \\[0.5em]
  &(k,l) \in R_b \text{ for some } a\neq b\Big\} \bigg| \,\,,
\end{align*}
where we consider the Euclidean norm. In other words the perimeter is the number of edges on the border between the regions of the segmentation - here we are thinking of an edge at the region's boundary as a pair of pixels on opposite sides of it, which have distance \(1\) in the Euclidean norm. If we assume the regions are all path connected when considering only paths that move horizontally or vertically in the image (i.e. not considering paths with segments of the type \((i,j),(i+1,j+1)\)), then these bordering edges completely define the segmentation. In any case, the perimeter gives a first relative estimate of the storage cost of the segmentation: in fact, though we haven't yet explicitly determined an efficient encoding of the perimeter edges, we assume that it will somehow encode the direction changes of the regions' border, such that e.g. a long straight border will be encoded as a sequence of \(0\) and thus will be efficiently compressible. 

\begin{center}
\begin{figure}[h]
\begin{subfigure}[b]{0.24\textwidth}
\includegraphics[scale=0.2]{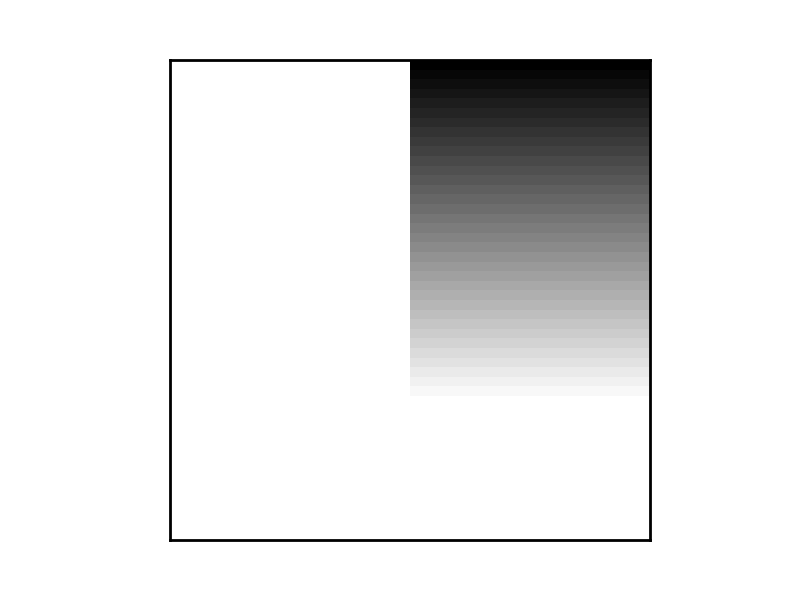}
\caption{}
\label{img:segm_ex}
\end{subfigure} 
\begin{subfigure}[b]{0.24\textwidth}
\includegraphics[scale=0.2]{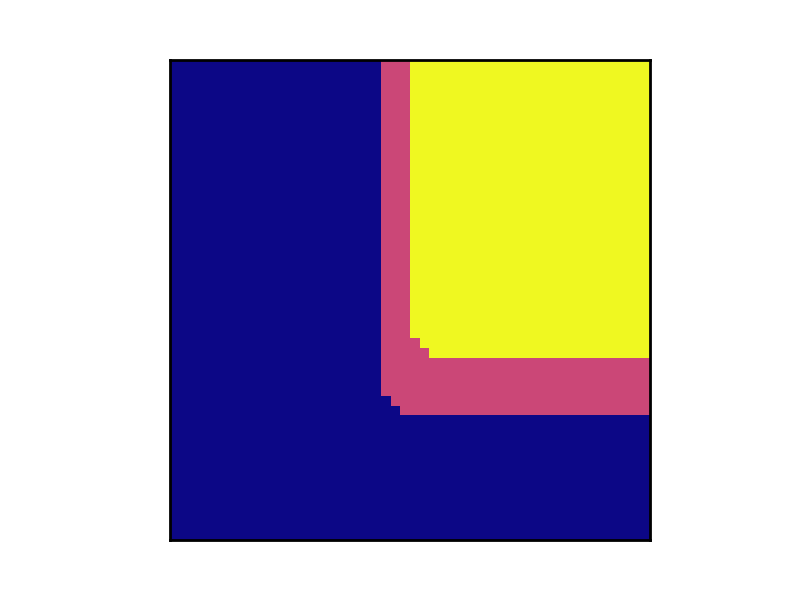}
\caption{}
\label{img:segm_ex-segm}
\end{subfigure} 
\begin{subfigure}[b]{0.24\textwidth}
\includegraphics[scale=0.2]{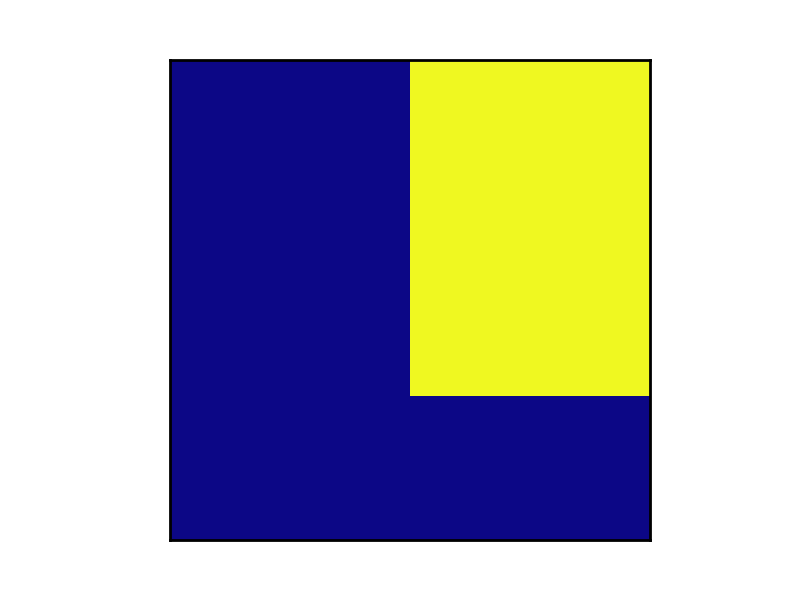}
\caption{}
\label{img:segm_ex-segm-sigma0}
\end{subfigure} 
\begin{subfigure}[b]{0.24\textwidth}
\includegraphics[scale=0.2]{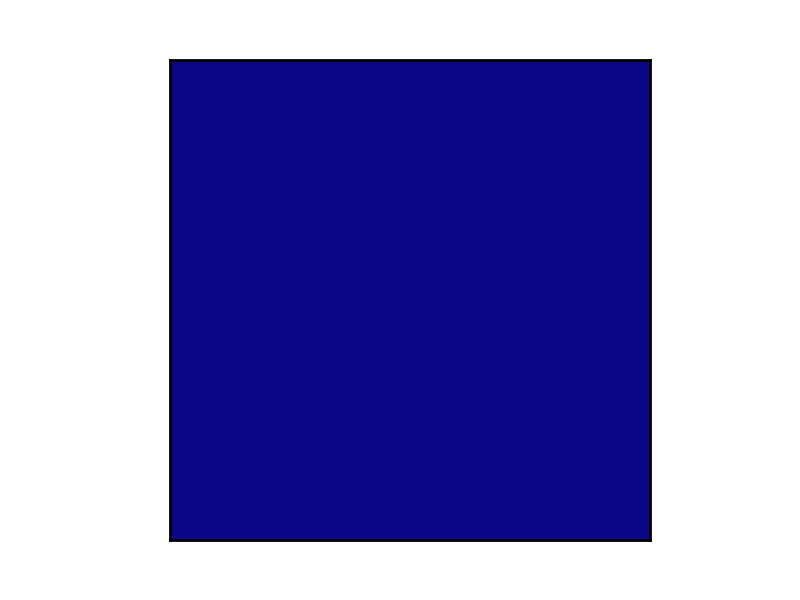}
\caption{}
\label{img:segm_ex-segm-bigscale}
\end{subfigure} 
\caption{(\subref{img:segm_ex}) Original image, (\subref{img:segm_ex-segm}) Felzenszwalb-Huttenlocher segmentation for some scale parameter \( k_0 \) and smoothing parameter \( \sigma>0 \), (\subref{img:segm_ex-segm-sigma0})  \( \sigma=0 \), and finally (\subref{img:segm_ex-segm-bigscale}) for \( \sigma=0 \) and scale parameter \( k_1 \gg k_0 \).}
\label{img:segm}
\end{figure}
\end{center}

\noindent Thus, having a small perimeter and regular borders are desirable characteristics of a segmentation for our method. A more fundamental goal is to produce regions such that, when its points are reordered according to the easy-path procedure, the gray-values considered in this order give a sequence with few jumps. A potentially tricky situation would be the image depicted in Figure  \ref{img:segm} (\subref{img:segm_ex}): segmentations (\subref{img:segm_ex-segm}) and especially (\subref{img:segm_ex-segm-sigma0}) are here clearly preferable to (\subref{img:segm_ex-segm-bigscale}), because the easy-path procedure applied to the latter one would pass many times through the central vertical line of the image and produce a signal with many jumps.

In our tests we always applied Felzenszwalb-Huttenlocher's segmentation proposed in \cite{felz-hutt} \footnote{more precisely its implementation in the \texttt{skimage} python library}, because of its very good performance to computation speed ratio. This method considers the graph \(G = (V,E)\), where the set of vertices are the pixels of the images and the edges connect each pixel to the 8 pixels forming its square neighborhood. The weights \(w(e)\) of the edges \(e\) are given by the difference in absolute value of the gray-values. Given any region (set of vertices) \(C\), its \emph{internal difference} is defined as
\begin{align}\label{eq:fhsegm.int}
C \subseteq V, \,\, \text{Int}(C) := \max{e\in \text{MST}(C,E)} w(e) \,\,,
\end{align}
where \(\text{MST}(C,E)\) is the minimum spanning tree of the subgraph with vertices \(C\) and all edges in \(E\) that are between vertices in \(C\), i.e. \((C,E\cap C\times C)\). This quantity measures the inner variance in gray-value, and tells us that there is a path between any two pixels in \(C\) using only edges of weight smaller than \(\text{Int}(C)\). Furthermore, given two regions \(C_1\) and \(C_2\), the authors in \cite{felz-hutt} define the \emph{difference} between the two regions as:
\begin{align}\label{eq:fhsegm.diff}
C_1,C_2 \subseteq V,  \,\, \text{Diff}(C_1,C_2) := \,\, \min{(v_i,v_j) \in C_1\times C_2 \cap E} \,\, w((v_i,v_j)) \,\, .
\end{align}
Finally, given two regions \(C_1\) and \(C_2\), they define a predicate, which evaluates to \texttt{True} if there is evidence of a boundary between the two regions and to \texttt{False} otherwise:
\begin{align}\label{eq:fhsegm.pred}
D(C_1,C_2) :=
\begin{cases}
   \texttt{True} & \text{ if }  \text{Diff}(C_1,C_2) > \min{}\{\text{I}_k(C_1),\text{I}_k(C_2)\}\\
   \texttt{False} & \text{ otherwise } 
\end{cases} \,\, ,
\end{align}
where 
\begin{align}\label{eq:fhsegm.intk}
\text{I}_k(C) := \text{Int}(C) + \frac{k}{|C|} 
\end{align}
is the \emph{scaled internal difference} of a region. The scaling parameter \(k\) is introduced to give a preference to regions of cardinality not too small with respect to \(k\). Note that instead of the term \(\frac{k}{|C|}\) any positive function \(\tau(C)\) could be used: the segmentation will then give a preference to regions \(C\) for which \(\tau(C)\) is big. In the future we wish to experiment with this function in order to optimize the regions for our method. One possible example would be to define \(\tau(C) = \frac{\text{area}(C)}{\text{perimeter}(C)}\), in order to obtain a segmentation with a smaller global perimeter and thus smaller storage cost.

\noindent The algorithm then starts with each pixel defined as its own region and then, considering each edge in increasing weight, evaluates the predicate for the two regions connected by the edge. It is proven to produce a segmentation which is neither too coarse nor too fine; see \cite{felz-hutt} for details. Prior to this procedure a Gaussian filter with variance \(\sigma\) is applied to smooth the image.

When comparing this segmentation method against the requirements of a good segmentation for our method, as described in the first part of this Section, we must make two remarks. Firstly, the regions produced are such that if we use the MST of that region to move between points, we will use only edges of low weight. But when we apply our method, we will be moving along the path generated by the easy-path procedure, and thus we will not have any guarantee of not obtaining a signal with jumps. Secondly, since in \eqref{eq:fhsegm.diff} the minimum edge weight is considered, the method can behave badly if there is a portion of the image similar to Figure \ref{img:segm}\subref{img:segm_ex} of size much smaller than the scale parameter \(k\). This could be avoided by substituting the minimum edge weight in \eqref{eq:fhsegm.diff} with the median edge weight or another quantile, but this is shown in \cite{felz-hutt} to make the segmentation problem NP-hard.

\subsection{Thresholding for a Region of Interest}
\label{sec:orgheadline8}
\label{orgtarget7}
\begin{figure}[h]
   \begin{subfigure}[b]{0.3\textwidth}
   \includegraphics[scale=0.2]{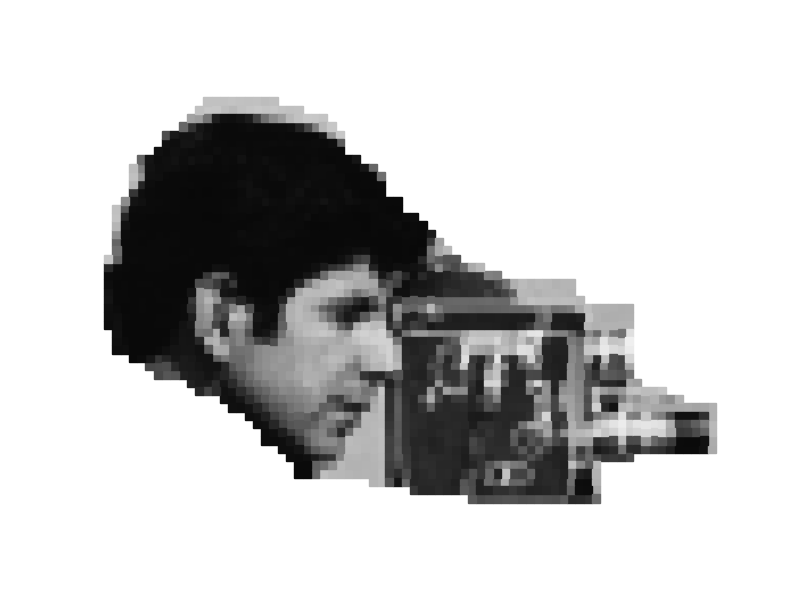}
\caption{}
 \label{fig:simpleroi:full}
   \end{subfigure} 
 \begin{subfigure}[b]{0.3\textwidth}
 \includegraphics[scale=0.2]{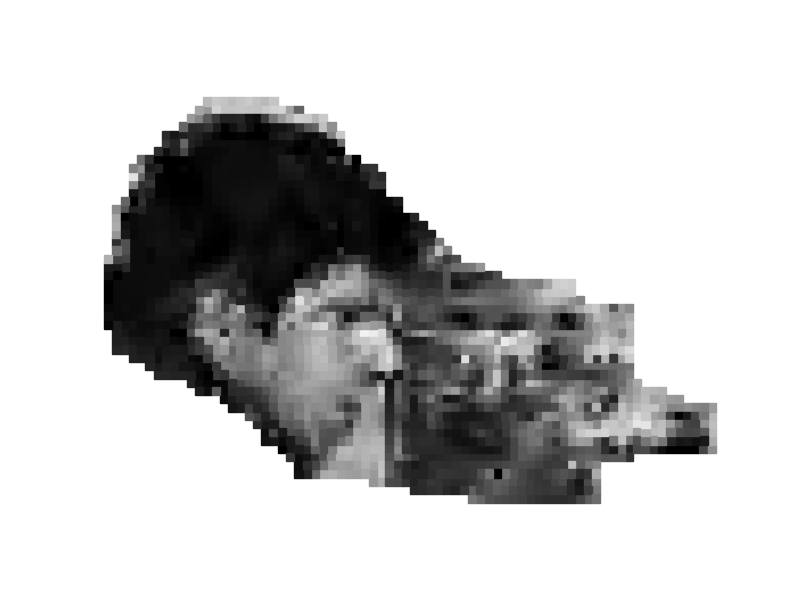}
\caption{}
 \label{fig:simpleroi:210}
   \end{subfigure}
   \begin{subfigure}[b]{0.3\textwidth}
   \includegraphics[scale=0.2]{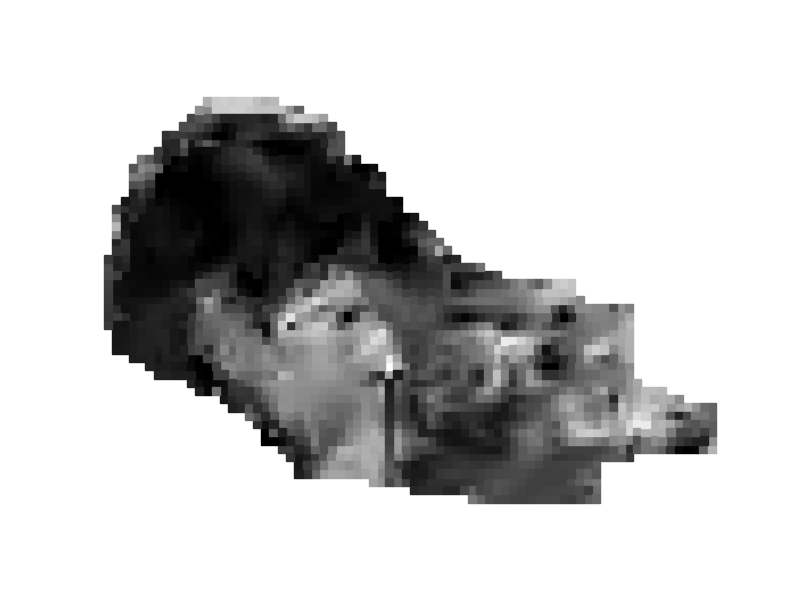}
\caption{}
 \label{fig:simpleroi:100}
   \end{subfigure}
   \caption{Encoding of a ROI for the cameraman image using the CDF \( 9/7 \) wavelet and \( 11 \) levels. (\subref{fig:simpleroi:full}): full quality with \( 2101 \) coefficients, (\subref{fig:simpleroi:210}): thresholded to \( 210 \) coefficients and (\subref{fig:simpleroi:100}): thresholded to \( 100 \) coefficients.}\label{fig:simpleroi}
   \end{figure}

Though we have supposed the index set to be that of a square image, i.e. \({I = \{ (i,j) \,\, | \,\, 0 \leq i,j \leq N-1 \}}\), there is nothing impeding us to relax this condition: the RBEPWT method can be applied without modification to images with arbitrary boundary shape. In particular, it can be used to encode only a region of interest (ROI) of a whole image - see Figure \ref{fig:simpleroi}.
One could then think of encoding with different qualities the ROI and the rest of the image, simply by viewing them as two separate images with non-regular boundaries. To this end however we can do something a little bit more clever, leveraging the fact that many coefficients of the full encoding of the image, especially at low levels, contain highly non-local information, shared by both the ROI and its complementary. This allows us fine control over the quality of the encoding for a region of arbitrary shape (and its complementary), something that is not possible with the classical \(2\)-dimensional wavelet transform.

For simplicity we will restrict ourselves to the case of Haar wavelets and we will suppose that the region of interest \(R\) is one of the regions found in the segmentation step; we will also suppose the image to be of size \(2^N\times 2^N\) and to encode using the maximum number of levels \(L=2\log_2 2^N = 2N\). The key observation is that each value of \(f^{L-1}\) will be determined only by two values of \(f^L\); this follows from \eqref{dwt:lev1} and the fact that the low and high pass Haar filters have support length \(2\). In order to determine all coefficients in the encoding necessary to reconstruct \(R\), it is useful to organize the coefficients in a graph structure, similarly as to how is done in \cite{gtbwt}, where a generalization of EPWT-like methods is proposed. There the authors define, for every level \(l\) of the transform, a node in the graph for each element in \(I^l\); edges are defined between a node from level \(l\) (i.e. a node associated to a point in \(I^l\)) to another at level \(l-1\) if the value of the latter is associated to a nonzero value of the filter in the sum defining the former. Thus this graph depends on the filter, and specifically only on the lengths of its support; in the Haar case, when the maximum number of levels is used, the graph is actually a full complete binary tree.

\begin{figure}[h]
 \begin{subfigure}[t]{0.48\textwidth}
\centering \includegraphics[scale=0.2]{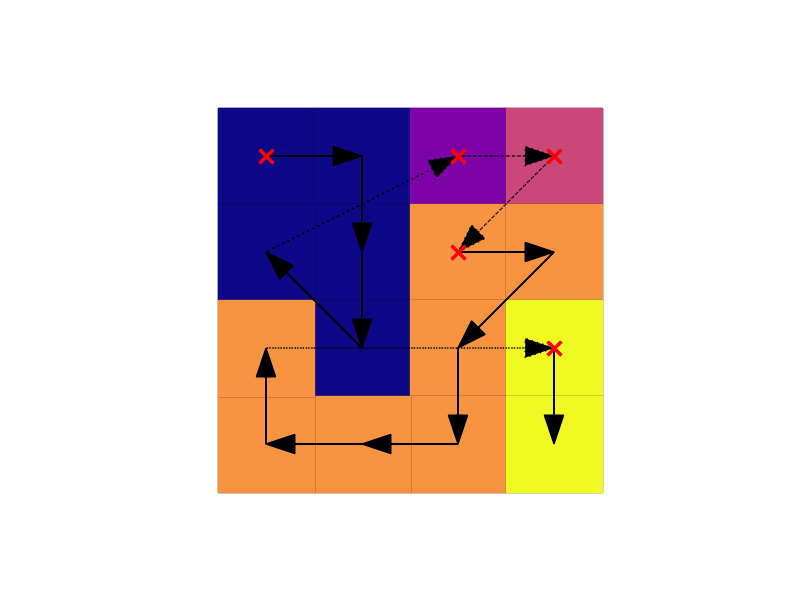}
\caption{}
\label{fig:coeffgraph:seg}
  \end{subfigure} 
 \begin{subfigure}[t]{0.48\textwidth}
\centering\includegraphics[scale=0.2]{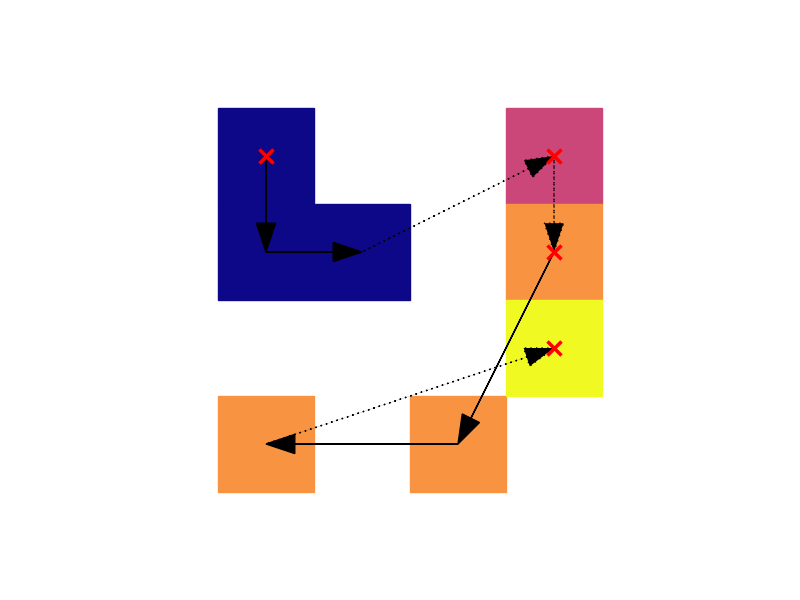}
\caption{}
\label{fig:coeffgraph:seglev2}
  \end{subfigure}
 \begin{subfigure}[t]{1\textwidth}
   \centering \includegraphics[scale=0.06]{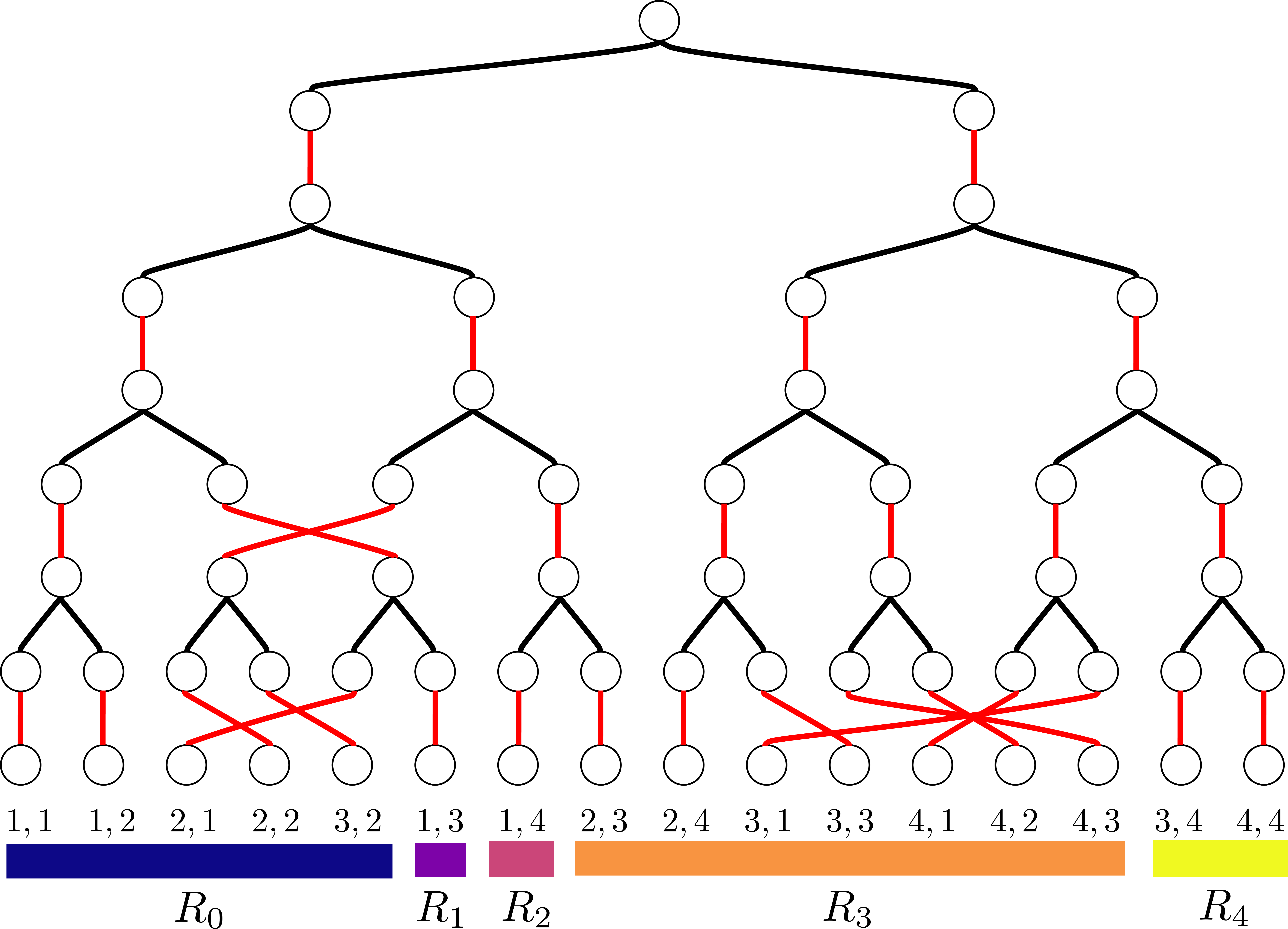}
\caption{}
\label{fig:coeffgraph:tree}
  \end{subfigure}
\caption{(\subref{fig:coeffgraph:seg}): A segmented \( 4\times 4 \) image with the path found by the easy path procedure at the fourth level, (\subref{fig:coeffgraph:seglev2}): the regions and path at the third level, (\subref{fig:coeffgraph:tree}): the coefficient tree associated to the RBEPWT with Haar wavelets and \( 4 \) levels. Here the implied order of the pixels is given by row-stacking (i.e. left to right and top to bottom). The leafs are denoted by their coordinates in the image and are grouped by region. The red edges represent the permutation given by the path finding procedure, while the black edges represent the convolution with the low and high pass filters.}\label{fig:coeffgraph}
\end{figure}

In order to visualize the permutation, we add to this construction for each level a set of nodes, again each representing a point in that level of the transform. We thus have for each point in \(I^l\) two nodes in the graph, and we can then view the permutation found at level \(l\) by the path-finding procedure of the RBEPWT as a bipartite subgraph. See Figure \ref{fig:coeffgraph} for an example.

Now, the points in \(R\) correspond to a set of leaf nodes in the tree; from the definition of the tree, it follows that only the RBEPWT coefficients corresponding to nodes ancestors of these leaf nodes are influenced by the values of the points in \(R\). The farther up the tree a node is, the larger is its zone of influence in the image - the root node is affected by all points in the image, but closer to the leafs the coefficients contain more localized information. For example in Figure \ref{fig:coeffgraph} \(5\)  coefficients (\(4\) detail and one approximation) are needed to perfectly reconstruct \(R_4\), and of these only one (the root node) is common to \(R_0\). If we wish to encode only region \(R\) then it suffices to follow the edges of the tree up until the root node, starting from the leafs associated to \(R\) and preserving the coefficients encountered in this visit while setting to \(0\) all the others. In this way all the information necessary to perfectly reconstruct \(R\) will be preserved alongside some information of the rest of the image; (see Figure \ref{fig:roi:full}). To encode with different qualities \(R\) and the rest of the image, we must divide the coefficients into two sets: those that are ancestors to points in \(R\) in the tree and those that are not. Then we threshold the coefficients, preserving a certain percentage of those in the first set and another percentage of those in the second but not in the first, always giving precedence to coefficients larger in absolute value (see Figure \ref{fig:roi:inout}). 

The general non-Haar case is more complicated: one point in \(I^l\) will influence \(\floor{s/2}\) coefficients at level \(l-1\), where \(s\) is the length of the filter and is usually greater than \(2\). This means the graph will not be a tree anymore, since there will be leafs with more than one parent; furthermore in the non-orthogonal case the reconstruction filters may have different lengths than the synthesis, thus making it not obvious from the graph representation which coefficients are needed for perfect reconstruction of the ROI.

Finally, this idea could be applied to the standard EPWT and tensor wavelet transform as well, however both cases would require more coefficients to encode a ROI compared to the RBEPWT case. For the EPWT, the permutation at each level is searched globally on the whole set of points and not for each region separately, thus the coefficients could potentially get mixed more, with points that are nearby in the image sharing less common coefficients in the graph. For the tensor wavelet transform, the number of coefficients at level \(l-1\) influenced by a single coefficient at level \(l\) is of the order of \((s/2)^2\) \footnote{supposing the low and high pass filter both have length \(s\)}, due to the nature of the transform that uses two-dimensional convolution.

\section{Numerical Experiments}
\label{sec:orgheadline10}
\label{orgtarget5}
\begin{figure}[p]
  \vspace{-3em}
\thisfloatpagestyle{empty}
\begin{subfigure}[b]{0.48\textwidth}
\includegraphics[scale=0.6]{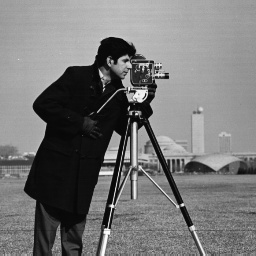}
        \caption{}
        \label{img:cameraman}
    \end{subfigure}%
\begin{subfigure}[b]{0.48\textwidth}
\hspace{0.1em} \includegraphics[scale=0.45]{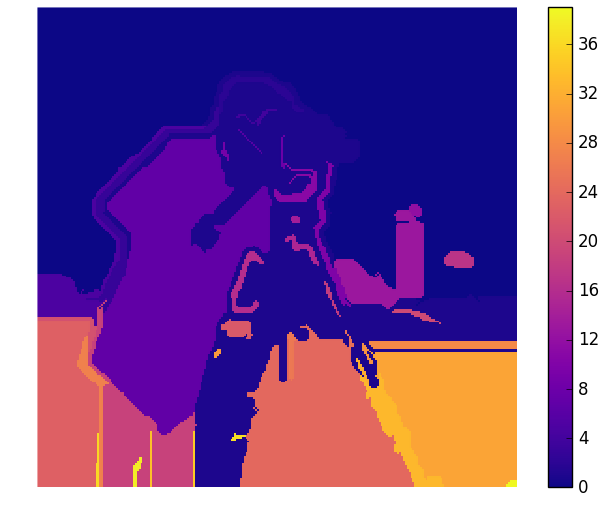}
\vspace{-2.2em}
        \caption{}
        \label{img:cameraman-segm}
    \end{subfigure}

\begin{subfigure}[b]{0.48\textwidth}
\includegraphics[scale=0.6]{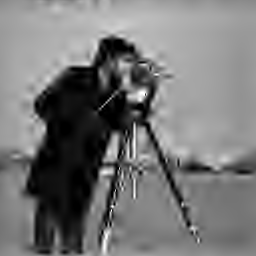}
        \caption{}
        \label{img:cameraman-tensor}
    \end{subfigure} \hspace*{\fill}
\begin{subfigure}[b]{0.48\textwidth}
\includegraphics[scale=0.6]{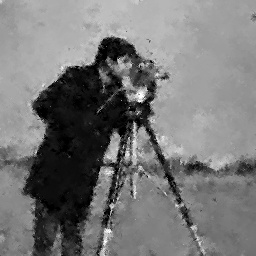}
        \caption{}
        \label{img:cameraman-epwt}
    \end{subfigure}

\begin{subfigure}[b]{0.48\textwidth}
\includegraphics[scale=0.6]{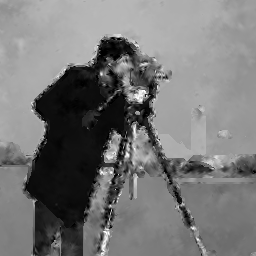}
        \caption{}
        \label{img:cameraman-easypath}
    \end{subfigure} \hspace*{\fill}
\begin{subfigure}[b]{0.48\textwidth}
\includegraphics[scale=0.6]{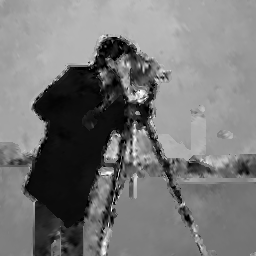}
        \caption{}
        \label{img:cameraman-gradpath}
    \end{subfigure}
\caption{(\subref{img:cameraman}): Original cameraman image. (\subref{img:cameraman-segm}): Segmentation obtained from the Felzenszwalb-Huttenlocher algorithm with scale parameter \( 200 \), \( \sigma = 2 \) and minimum scale \( 10 \). (\subref{img:cameraman-tensor})-(\subref{img:cameraman-gradpath}): Image compressed using 512 coefficients and the CDF 7/9 filter: (\subref{img:cameraman-tensor}) the classical 2D tensor wavelet transform, (\subref{img:cameraman-epwt}) EPWT, (\subref{img:cameraman-easypath}) RBEPWT with easy-path and (\subref{img:cameraman-gradpath}) RBEPWT with grad-path transforms.}
\label{img:comparison.cameraman}
\end{figure}

\begin{figure}[p]
  \vspace{-3em}
  \thisfloatpagestyle{empty}
\begin{subfigure}[b]{0.48\textwidth}
\includegraphics[scale=0.6]{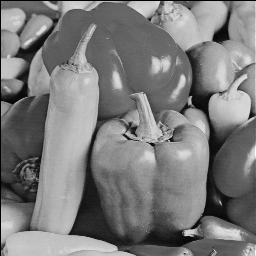}
        \caption{}
        \label{img:peppers}
    \end{subfigure}%
\begin{subfigure}[b]{0.48\textwidth}
\hspace{0.1em} \includegraphics[scale=0.45]{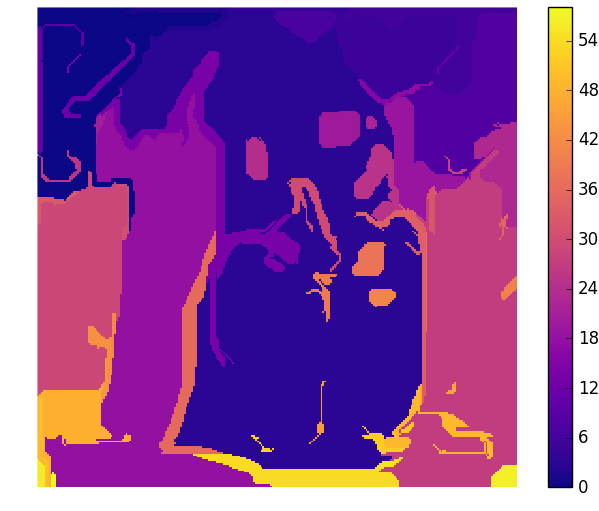}
\vspace{-2.2em}
        \caption{}
        \label{img:peppers-segm}
    \end{subfigure}

\begin{subfigure}[b]{0.48\textwidth}
\includegraphics[scale=0.6]{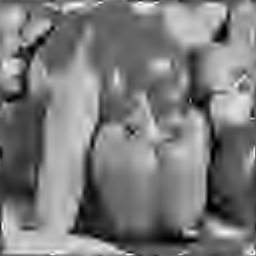}
        \caption{}
        \label{img:peppers-tensor}
    \end{subfigure} \hspace*{\fill}
\begin{subfigure}[b]{0.48\textwidth}
\includegraphics[scale=0.6]{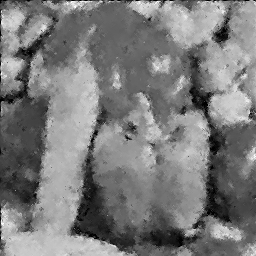}
        \caption{}
        \label{img:peppers-epwt}
    \end{subfigure}

\begin{subfigure}[b]{0.48\textwidth}
\includegraphics[scale=0.6]{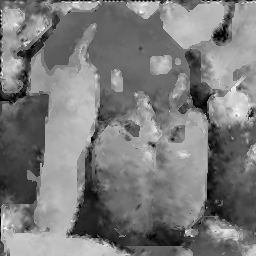}
        \caption{}
        \label{img:peppers-easypath}
    \end{subfigure} \hspace*{\fill}
\begin{subfigure}[b]{0.48\textwidth}
\includegraphics[scale=0.6]{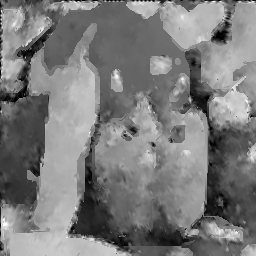}
        \caption{}
        \label{img:peppers-gradpath}
    \end{subfigure}
\caption{(\subref{img:peppers}): Original peppers image. (\subref{img:peppers-segm}): Segmentation obtained from the Felzenszwalb-Huttenlocher algorithm with scale parameter \( 200 \), \( \sigma = 2 \) and minimum scale \( 10 \). (\subref{img:peppers-tensor})-(\subref{img:peppers-gradpath}): Image compressed using 512 coefficients and the CDF 7/9 filter: (\subref{img:peppers-tensor}) the classical 2D tensor wavelet transform, (\subref{img:peppers-epwt}) EPWT, (\subref{img:peppers-easypath}) RBEPWT with easy-path and (\subref{img:peppers-gradpath}) RBEPWT with grad-path transforms.}
\label{img:comparison.peppers}
\end{figure}

\begin{figure}[p]
  \vspace{-3em}
  \thisfloatpagestyle{empty}
\begin{subfigure}[b]{0.48\textwidth}
\includegraphics[scale=0.6]{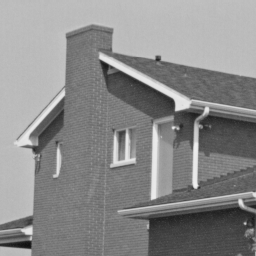}
        \caption{}
        \label{img:house}
    \end{subfigure}%
\begin{subfigure}[b]{0.48\textwidth}
\hspace{0.1em} \includegraphics[scale=0.45]{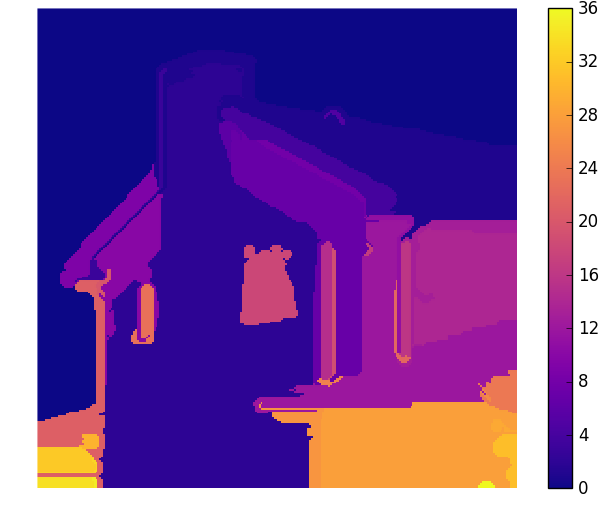}
\vspace{-2.2em}
        \caption{}
        \label{img:house-segm}
    \end{subfigure}

\begin{subfigure}[b]{0.48\textwidth}
\includegraphics[scale=0.6]{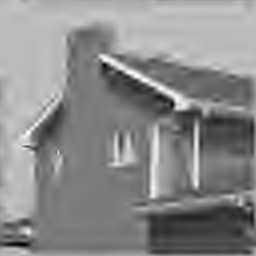}
        \caption{}
        \label{img:house-tensor}
    \end{subfigure} \hspace*{\fill}
\begin{subfigure}[b]{0.48\textwidth}
\includegraphics[scale=0.6]{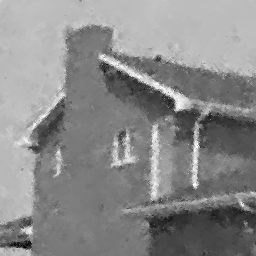}
        \caption{}
        \label{img:house-epwt}
    \end{subfigure}

\begin{subfigure}[b]{0.48\textwidth}
\includegraphics[scale=0.6]{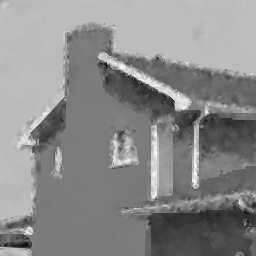}
        \caption{}
        \label{img:house-easypath}
    \end{subfigure} \hspace*{\fill}
\begin{subfigure}[b]{0.48\textwidth}
\includegraphics[scale=0.6]{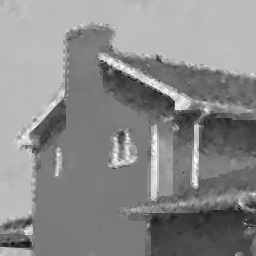}
        \caption{}
        \label{img:house-gradpath}
    \end{subfigure}
\caption{(\subref{img:house}): Original house image. (\subref{img:house-segm}): Segmentation obtained from the Felzenszwalb-Huttenlocher algorithm with scale parameter \( 200 \), \( \sigma = 2 \) and minimum scale \( 10 \). (\subref{img:house-tensor})-(\subref{img:house-gradpath}): Image compressed using 512 coefficients and the CDF 7/9 filter: (\subref{img:house-tensor}) the classical 2D tensor wavelet transform, (\subref{img:house-epwt}) EPWT, (\subref{img:house-easypath}) RBEPWT with easy-path and (\subref{img:house-gradpath}) RBEPWT with grad-path transforms.}
\label{img:comparison.house}
\end{figure}

We present here some numerical results, which were all obtained using the python code avaiable at the software page of \href{http://na.math.uni-goettingen.de/}{\url{http://na.math.uni-goettingen.de/}} \footnote{also avaiable at  \url{https://github.com/nareto/rbepwt}}. We used the biorthogonal Cohen-Daubechies-Feauveau wavelets (known also as CDF \(9/7\)) because, analogously to other natural images compression methods, they showed very good performance in our method compared to other wavelets. In Figures \ref{img:comparison.cameraman}-\ref{img:comparison.house} our method is compared to the classical two-dimensional tensor wavelet transform and the EPWT. It is interesting to note how thresholding the coefficients introduces different types of distortion in the different transforms: the 2D wavelet transform blurs the image, while the EPWT and RBEPWT methods preserve edges much better, while introducing a higher frequency noise. In particular, as should be expected, the RBEPWT preserves the segmentation information: for example in the cameraman image, the skyscraper to the right of the camera tripod can be seen only in the RBEPWT images, because it was identified as a separate region in the segmentation step. In the house image the texture of the bricks is completely gone, but the borders are still perfectly visible - thus our method is particularly well-suited for cartoon-like images.

\begin{figure}[htbp]
    \vspace{-3em}
\thisfloatpagestyle{empty}
\begin{subfigure}[b]{\textwidth}
\centerline{\includegraphics[scale=0.4]{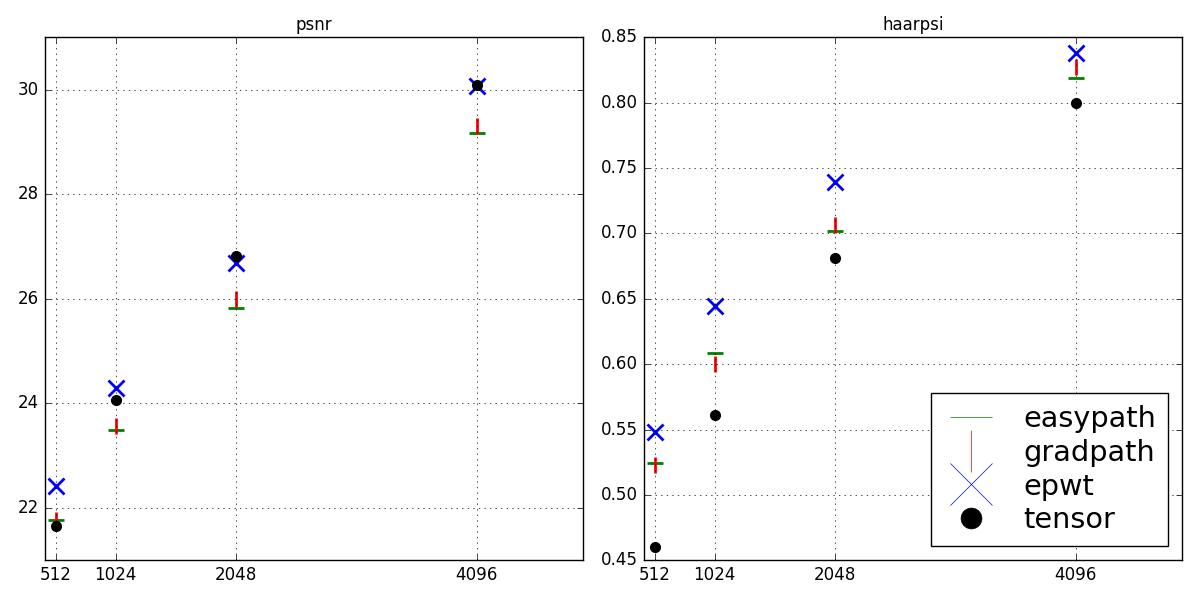}}
\caption{}
\label{img:errors.cameraman}
\end{subfigure}

\begin{subfigure}[b]{\textwidth}
\centerline{\includegraphics[scale=0.4]{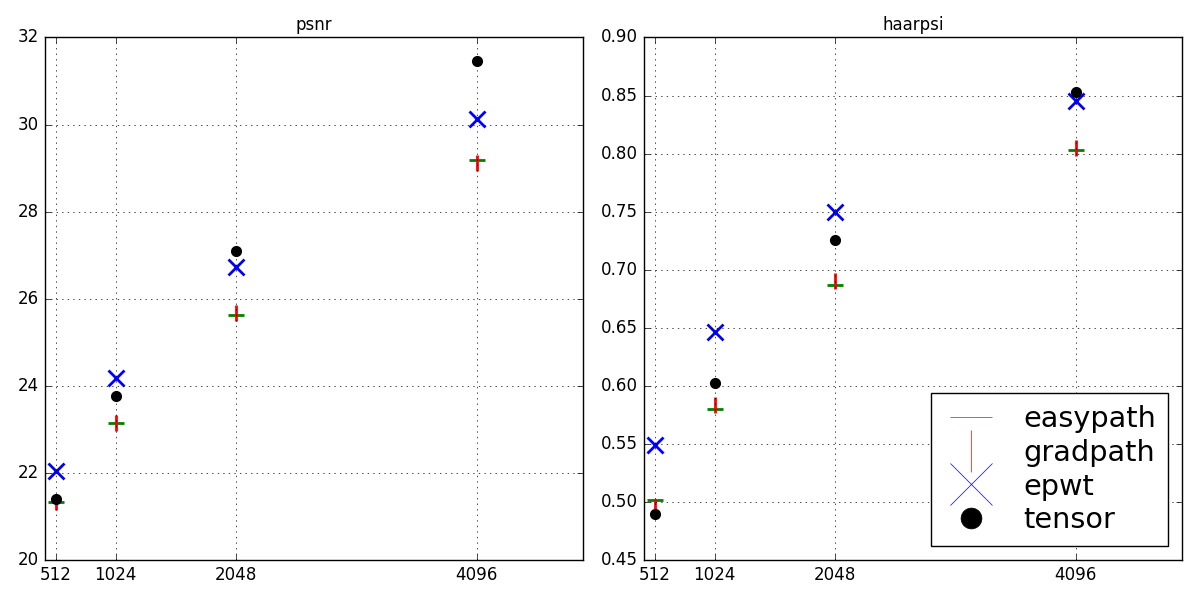}}
\caption{}
\label{img:errors.peppers}
\end{subfigure}

\begin{subfigure}[b]{\textwidth}
\centerline{\includegraphics[scale=0.4]{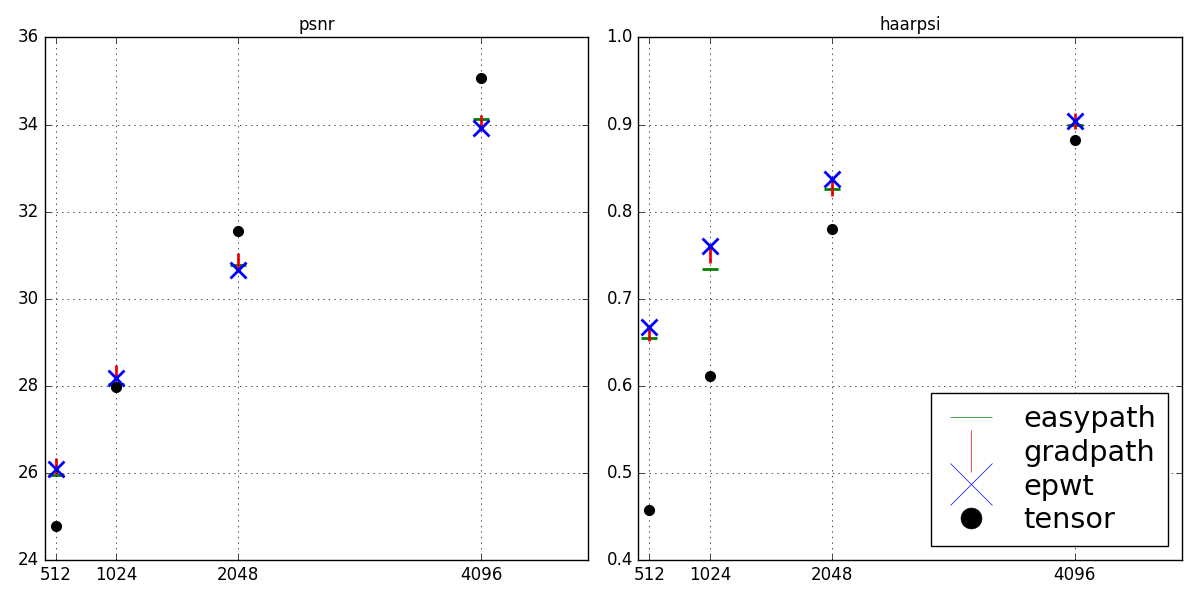}}
\caption{}
\label{img:errors.house}
\end{subfigure}

\caption{\label{fig:errors}PSNR and HaarPSI values for the (\subref{img:errors.cameraman}) cameraman, (\subref{img:errors.peppers}) peppers and (\subref{img:errors.house}) house image, with \( 512, 1024, 2048 \) and \( 4096 \) coefficients}
\label{img:errors}
\end{figure}

\noindent In order to quantify the quality of our method, we computed the peak signal to noise ratio (PSNR)
\begin{align*}
\text{PSNR}(f,\tilde{f}) :=  20 \log_2{\frac{255}{\norm{f - \tilde{f}}_2}} \,\, ,
\end{align*}
where here \(f\) is the matrix corresponding to the original image and \(\tilde{f}\) to the reconstruction using only the most significant RBEPWT  coefficients. However, in recent years there has been a lot of research to find a better quality measurement index (for example see \cite{ssim} and \cite{vsi}). Here by better we mean that it correlates better to scores given by humans to different degraded versions of the same image, such as those presented in \cite{TID2013} or \cite{CSIQ}. In \cite{haarpsi}, the authors show that on average their proposed HaarPSI index has the best correlation to such databases. Thus, in Figure \ref{img:errors}, we show the PSNR (for comparison with other methods, since despite its shortcomings it is still widely used) and the HaarPSI values for the cameraman, peppers and house image. These plots show that our method, compared to EPWT, performs very good on the house image, which is almost cartoon-like, and a little worse on the cameraman and especially peppers image.

\noindent It should be noted here that a better comparison would be obtained by comparing the quality of the thresholded images with different transforms having the same storage cost (and not the same number of coefficients). However at this point it is unclear to us how the segmentation can be efficiently encoded.

\begin{figure}[h!]
 \begin{subfigure}[b]{0.48\textwidth}
 \includegraphics[scale=0.45]{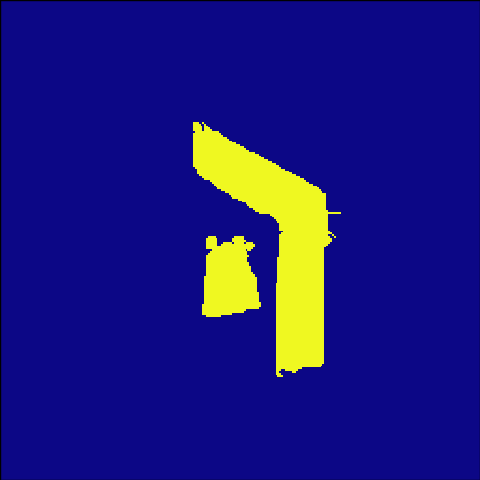}
 \caption{}
\label{fig:roi:seg}
 \end{subfigure}
 \begin{subfigure}[b]{0.48\textwidth}
 \includegraphics[scale=0.6]{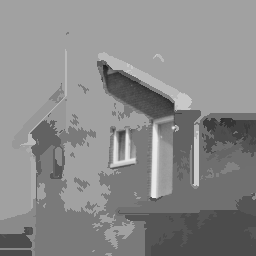}
 \caption{}
\label{fig:roi:full}
\end{subfigure} \hspace*{\fill}

 \begin{subfigure}[b]{0.48\textwidth}
 \includegraphics[scale=0.6]{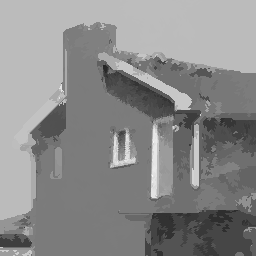}
 \caption{}
\label{fig:roi:inout}
 \end{subfigure}
 \begin{subfigure}[b]{0.48\textwidth}
 \includegraphics[scale=0.6]{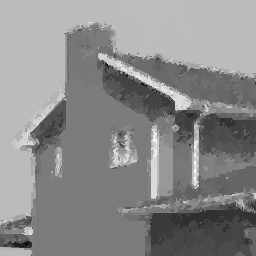}
 \caption{}
\label{fig:roi:flat}
\end{subfigure} \hspace*{\fill}

\caption{(\subref{fig:roi:seg}): the selected ROI for the house image, roughly corresponding to the window, the light part of the wall and the area under the roof; (\subref{fig:roi:full}): decoded image where all and only the coefficients ancestors to the ROI where preserved (5042 non-zero coefficients); (\subref{fig:roi:inout}): decoded image where \( 10\% \) of coefficients ancestors to the ROI  and \( 0.1\% \) not ancestors were preserved, for a total of \( 551 \) non-zero coefficients; (\subref{fig:roi:flat}): decoded image with standard thresholding with \( 551 \) non-zero coefficients (inserted here for comparison with (\subref{fig:roi:inout}))}
\label{fig:roi}
 \end{figure}

\noindent In Figure \ref{fig:roi} we show an example of the ROI based thresholding described in section \ref{orgtarget7}. 

\begin{center}
  \begin{figure}[p]
      \vspace{-3em}
\thisfloatpagestyle{empty}
\begin{subfigure}[b]{0.48\textwidth}
\includegraphics[scale=0.6]{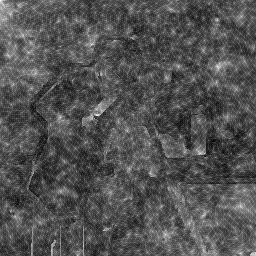}
        \caption{}
        \label{img:cameraman.rbepwt.basis:17.0}
    \end{subfigure} \hspace*{\fill}
\begin{subfigure}[b]{0.48\textwidth}
\includegraphics[scale=0.6]{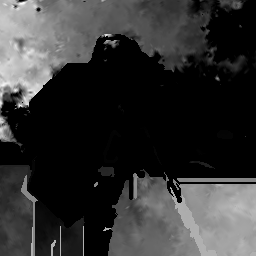}
        \caption{}
        \label{img:cameraman.rbepwt.basis:16.0}
    \end{subfigure}

\begin{subfigure}[b]{0.48\textwidth}
\includegraphics[scale=0.6]{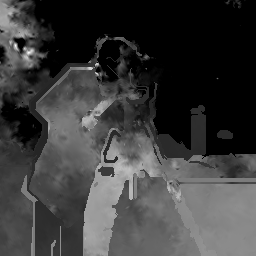}
        \caption{}
        \label{img:cameraman.rbepwt.basis:15.0}
    \end{subfigure} \hspace*{\fill}
\begin{subfigure}[b]{0.48\textwidth}
\includegraphics[scale=0.6]{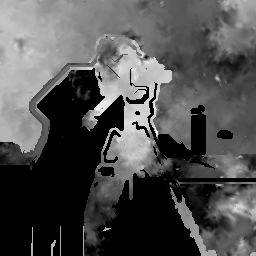}
\caption{}
        \label{img:cameraman.rbepwt.basis:15.1}
    \end{subfigure}

\begin{subfigure}[b]{0.48\textwidth}
\includegraphics[scale=0.6]{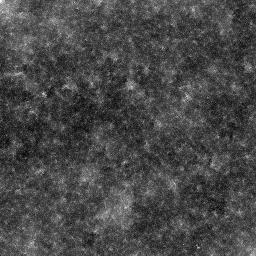}
        \caption{}
        \label{img:cameraman.epwt.basis:17.0}
    \end{subfigure} \hspace*{\fill}
\begin{subfigure}[b]{0.48\textwidth}
\includegraphics[scale=0.6]{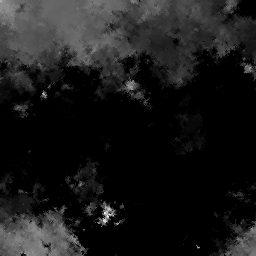}
        \caption{}
        \label{img:cameraman.epwt.basis:16.0}
    \end{subfigure}
\caption{Basis elements for the cameraman image, encoded with 16 levels. (\subref{img:cameraman.rbepwt.basis:17.0})-(\subref{img:cameraman.rbepwt.basis:16.0}): approximation and detail coefficients at the first level of the RBEPWT; (\subref{img:cameraman.rbepwt.basis:15.0})-(\subref{img:cameraman.rbepwt.basis:15.1}) the two detail coefficients at the second level of the RBEPWT; (\subref{img:cameraman.epwt.basis:17.0})-(\subref{img:cameraman.epwt.basis:16.0}) approximation and detail coefficients at the first level of the EPWT.}
\label{img:cameramanbasis}
\end{figure}
\end{center}

\noindent Finally, we observe that by setting one coefficient to \(1\) and the rest to \(0\) and then decoding, it is possible to view the elements of the adaptive basis produced by our method. In Figure \ref{img:cameramanbasis} some of such elements are shown for the cameraman image, both for the RBEPWT and the EPWT. As should be expected, the basis for our method clearly preserves the segmentation structure, while the basis for the EPWT appears to be totally unstructured.

\section{Final Remarks}
\label{sec:orgheadline11}
We introduced a new method for compression of natural images, which applies an image segmentation method to the image before applying an EPWT-like method. We defined two region path-finding procedures, which are deterministic and do not depend on the single gray values of the pixels in the region. We used the defined transform to compute a lossy compressed version of the original image, obtained by setting to \(0\) all but the greatest coefficients and applying the decoding procedure. We commented on which are desirable characteristics for the segmentation step of our method, namely producing regions that enable paths generated by the easy-path procedure with few jumps and with small and regular perimeters. Finally we showed how the RBEPWT can be used to encode with different quality a specific region of interest from the rest of the image.

\noindent In the future we wish to study more closely the segmentation step and possibly develop an ad-hoc procedure for our method.

\section*{Acknowledgments}
\label{sec:orgheadline12}
The author gratefully acknowledges the funding of this work by the DFG in the framework of the GRK 2088 and heartfully thanks Gerlind Plonka whose guidance and insight have been fundamental for the completion of this work.

\bibliographystyle{plain}
\bibliography{biblio}{}
\end{document}